\documentclass[
    ,final            
  ]
  {aipproc}

\layoutstyle{8x11single}

\newcommand{\h}[1]{H~{\sc #1}}
\newcommand{\gray}[1]{$\gamma$-ray{#1}}
\newcommand{\cray}[1]{CR{#1}}
\newcommand{\pubjournal}[6] {#1, #2 {\bf #3}, #4 (#5).}

\newcommand{\aap}{{\it Astron. Astrophys.}}
\newcommand{\apj}{{\it ApJ}}
\newcommand{\adv}{{\it Adv. Space Res.}}

\begin{document}

\title{Developing the Galactic diffuse emission model for the 
GLAST Large Area Telescope}


\author{Igor V. Moskalenko}{
  address={Hansen Experimental Physics Laboratory,
  Stanford University, Stanford, CA 94305},
  altaddress={Kavli Institute for Particle Astrophysics and Cosmology,
    Stanford University, Stanford, CA 94309}
}

\author{Andrew W. Strong}{
  address={Max-Plank-Institut f\"ur extraterrestrische Physik, 
  Postfach 1312, D-85741 Garching, Germany}
}

\author{Seth W. Digel}{
  address={Stanford Linear Accelerator Center,
  2575 Sand Hill Rd, Menlo Park, CA 94025},
  altaddress={Kavli Institute for Particle Astrophysics and Cosmology,
    Stanford University, Stanford, CA 94309}
}

\author{Troy A. Porter}{
  address={Santa Cruz Institute for Particle Physics, 
  University of California, Santa Cruz, CA 95064}
}

\begin{abstract}
Diffuse emission is produced in energetic cosmic ray (\cray{}) interactions,
mainly protons and electrons, with the interstellar gas and radiation
field and contains the information about particle spectra in distant
regions of the Galaxy. It may also contain information about exotic
processes such as dark matter annihilation, black hole evaporation
etc. A model of the diffuse emission is important for determination of the
source positions and spectra. Calculation of the Galactic diffuse
continuum \gray{} emission requires a model for \cray{}
propagation as the first step. Such a model is based on theory of
particle transport in the interstellar medium as well as on many kinds
of data provided by different experiments in Astrophysics and Particle
and Nuclear Physics. Such data include: secondary particle and
isotopic production cross sections, total interaction nuclear cross
sections and lifetimes of radioactive species, gas mass calibrations and gas
distribution in the Galaxy (H$_2$, \h{i}, \h{ii}), interstellar radiation
field, \cray{} source distribution and
particle spectra at the sources, magnetic field, energy losses,
\gray{} and synchrotron production mechanisms, and many other
issues. We are continuously improving the GALPROP model and the code
to keep up with a flow of new data. Improvement in any field may
affect the Galactic diffuse continuum \gray{} emission model used as
a background model by the GLAST LAT instrument. Here we report about
the latest improvements of the GALPROP and the diffuse emission model.

\end{abstract}

\maketitle


\section{Discussion and Results}

We give a very brief summary of GALPROP; 
for details we refer 
to the relevant papers
\cite{1998ApJ...509..212S}-%
\cite{2006ApJ...642..902P}
and a dedicated website. 
The propagation equation is solved numerically on a spatial grid,
either in 2D with cylindrical symmetry in the Galaxy or in full 3D.
The boundaries of the model in radius and height, and the grid
spacing, are user-definable. Parameters for all
processes in the propagation equation can be controlled on input.  The
distribution of CR sources can be freely chosen, typically to
represent supernova remnants.  Source spectral shape and isotopic composition
(relative to protons) are input parameters.
Cross-sections are based on extensive compilations and
parameterizations \cite{2004AdSpR..34.1288M}.
The numerical solution is evolved forward in time until a steady-state is reached; 
a time-dependent solution is also an option. Starting with
the heaviest  primary nucleus considered (e.g., $^{64}$Ni) the
propagation solution is used to compute the source term for its
spallation products, which are then propagated in turn, and so on down
to protons, secondary electrons and positrons, and antiprotons.  In
this way secondaries, tertiaries, etc., are included.  Primary electrons are
treated separately.  The proton, helium, and electron spectra are 
normalized to data; all other isotopes are determined
by the source composition and propagation.
\gray{s} and synchrotron emission are computed  using interstellar gas data 
(for pion-decay and bremsstrahlung) and the interstellar radiation field 
(ISRF) model (for inverse Compton).
The computing resources required by GALPROP are moderate by current
standards.  

Recent extensions to GALPROP include
\begin{itemize}

\item
new detailed calculation of the ISRF \cite{isrf,isrf1}

\item
proper implementation of the anisotropic inverse Compton scattering
using new ISRF (Figure 1, left)

\item 
interstellar gas distributions based on current HI and CO surveys 
\cite{HIsurvey,COsurvey}

\item 
new parameterization of the $\pi^0$ production in $pp$-collisions
\cite{Tune} which includes the diffraction dissociation

\item 
non-linear MHD wave -- particle interactions (wave damping)
\cite{2006ApJ...642..902P} are included as an option

\item
the kinetic energy range is now extended down to $\sim$1 keV

\item
the \gray{} calculations extend from keV to tens of TeV (e.g., Figure 1, right), 
and produce full sky maps as a function of energy; 
the output is in the FITS-format

\item
gas mass calibration ($X_{\rm CO}$-factors) which can vary with position

\item
a dark matter package to allow for propagation of the WIMP annihilation 
products and calculation of the corresponding synchrotron and \gray{} skymaps

\item
GALPROP--DarkSUSY interface (together with T. Baltz) 
will become publicly available soon

\item
a dedicated website has been developed (\url{http://galprop.stanford.edu}) 

\end{itemize}

\begin{figure}
  \includegraphics[height=.27\textheight]{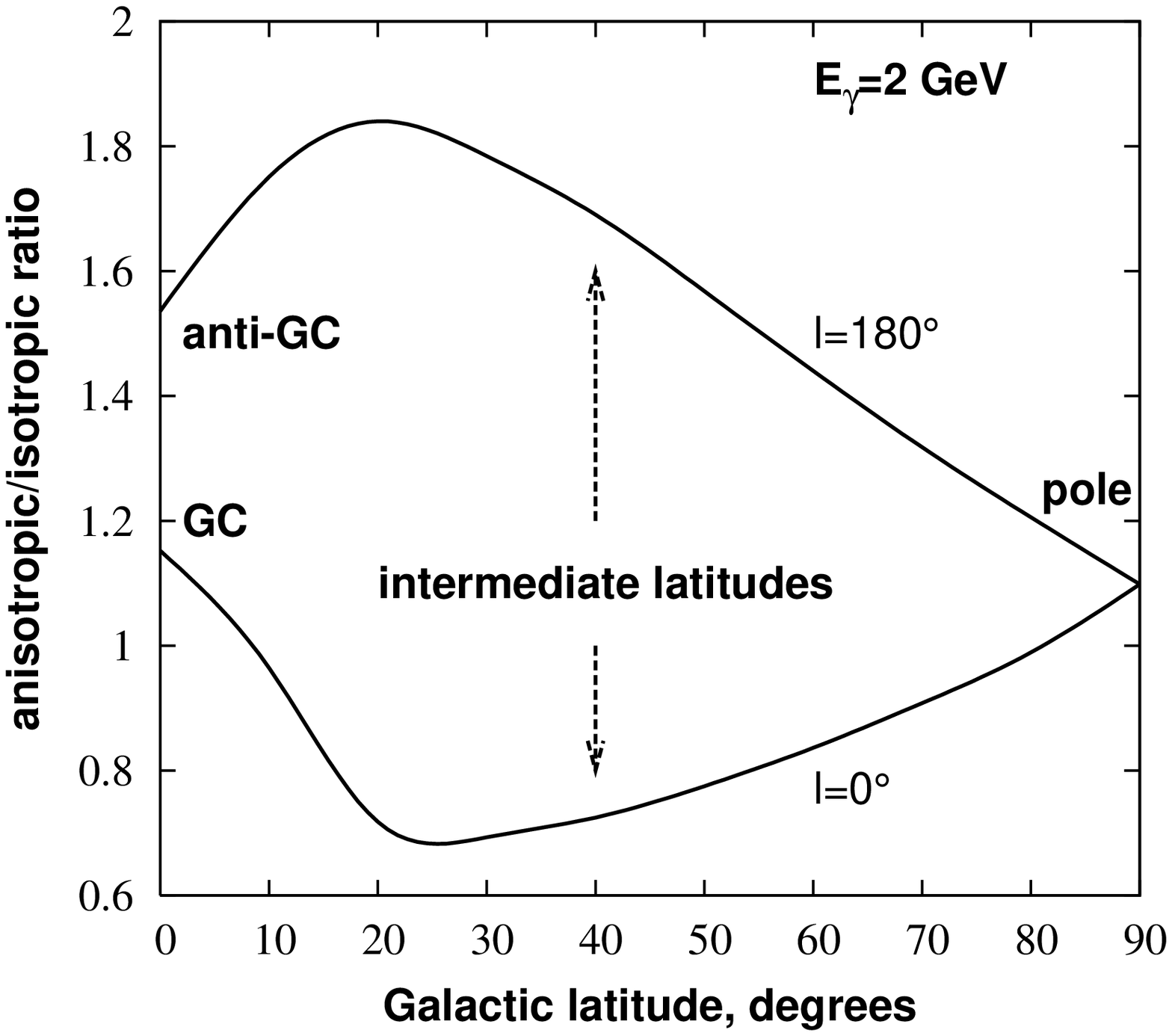}
  \includegraphics[height=.27\textheight]{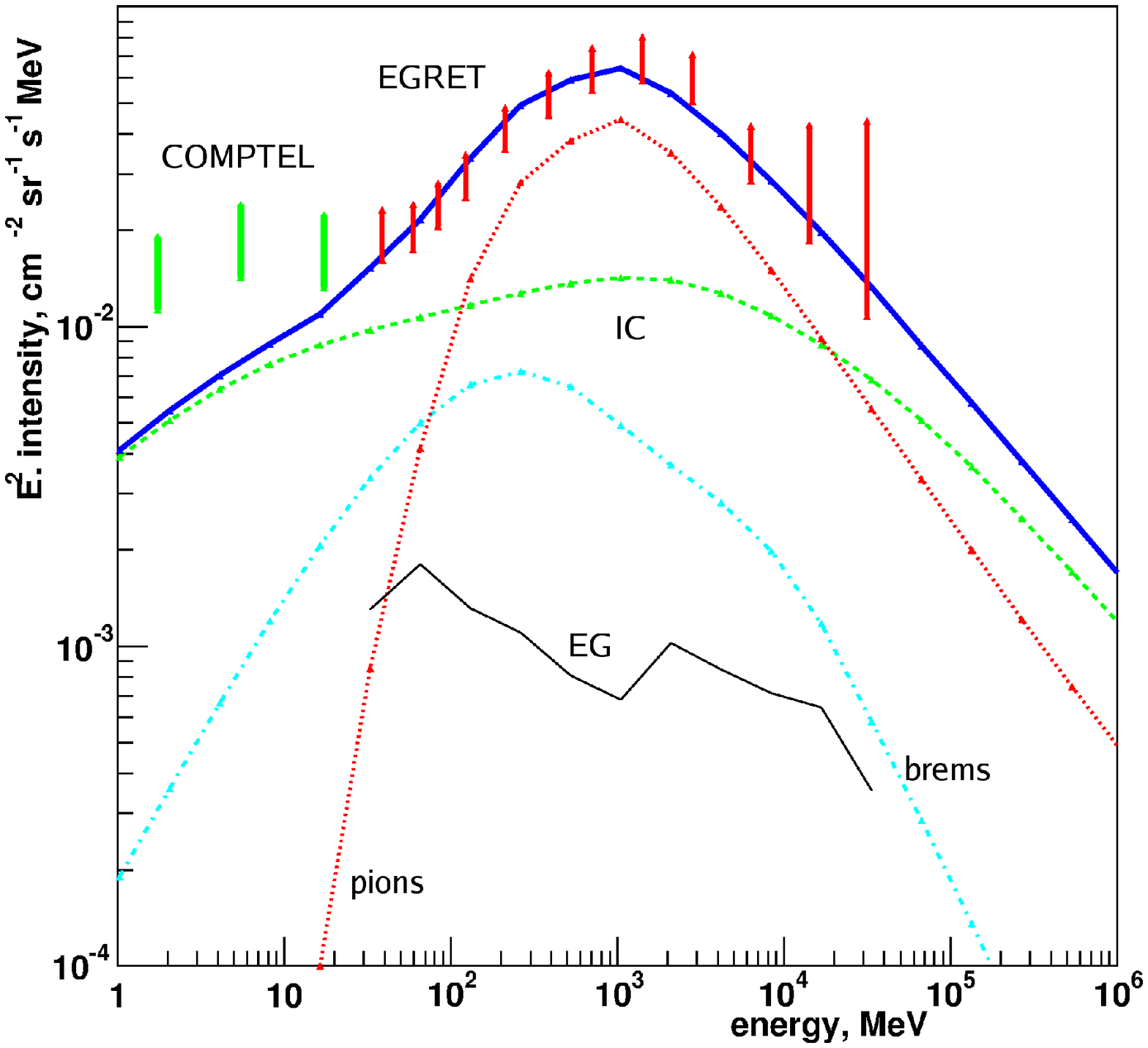}
  \caption{{\it Left:} The ratio of anisotropic IC to isotropic IC for Galactic 
longitudes $l=0^\circ$ and 180$^\circ$ vs.\ Galactic latitude.
{\it Right:} \gray{} spectrum of inner Galaxy ($330^\circ<l<30^\circ, |b|<5^\circ$) 
for an optimized model. Vertical bars: COMPTEL and EGRET data,
heavy solid line: total calculated flux.
This is an update of the spectrum shown in \cite{2004ApJ...613..962S}.}
\end{figure}

The GALPROP code
\cite{1998ApJ...509..212S}
was created with the following aims: (i) to enable simultaneous
predictions of all relevant observations including \cray{} nuclei,
electrons and positrons, \gray{s} and synchrotron radiation, (ii) to
overcome the limitations of analytical and semi-analytical methods,
taking advantage of advances in computing power, as \cray{}, \gray{}
and other data become more accurate, (iii)  to incorporate current
information on Galactic structure and source distributions, (iv) to
provide a publicly-available code as a basis for further expansion.
The first point is the most important: all data relating 
to the same system, the Galaxy, must have an internal consistency.
For example, one cannot
allow a model which fits secondary/primary ratios while not fitting
\gray{s} or not being compatible with the known interstellar gas
distribution.  There are many simultaneous constraints, and to find
one model satisfying all of them is a challenge,  which in fact has
not been met up to now.  Upcoming missions will
benefit: GALPROP has been adopted as the standard for diffuse
Galactic \gray{} emission  for NASA's GLAST \gray{} observatory,
and is also made use of by the ACE, AMS, HEAT and Pamela
collaborations.



IVM is supported in part by NASA APRA grant,
TAP is supported in part by the US Department of Energy.



\bibliographystyle{aipproc}   

\end{document}